\documentclass[prd,twocolumn,showpacs,preprintnumbers,nofootinbib,amsmath,amssymb]{revtex4}

\usepackage{graphicx}% Include figure files
\usepackage{dcolumn}% Align table columns on decimal point
\usepackage{bm}% bold math
\usepackage{amsmath}

\newcommand{\beq}{\begin{eqnarray}}
\newcommand{\eeq}{\end{eqnarray}}

\newcommand{\pardis}{\langle \mu \rangle}

\newcommand{\ie}{{\it i.e.\ }}

\newcommand{\real}{{\sf I}\kern-.12em{\sf R}}
\newcommand{\comp}{{\sf I}\kern-.50em{\sf C}}
\newcommand{\unity}{{\sf I}\kern-.54em{\sf 1}}

\newcommand{\tmtextrm}[1]{{\rmfamily{#1}}}

\def\spose#1{\hbox to 0pt{#1\hss}}
\def\ltapprox{\mathrel{\spose{\lower 3pt\hbox{$\mathchar"218$}}
 \raise 2.0pt\hbox{$\mathchar"13C$}}}

\begin{document}

\preprint{GEF-TH 15/06}
\preprint{UB-ECM-PF-06/25}

\title{Direct numerical computation of disorder parameters 
}
\author{Massimo D'Elia$^{1}$ and Luca Tagliacozzo$^{2}$}
%\email{massimo.delia@ge.infn.it; luca@ecm.ub.es}
\affiliation{$^1$  Dipartimento di Fisica dell'Universit\`a di Genova and INFN, Sezione di Genova, Via Dodecaneso 33, I-16146 Genova, Italy}
\affiliation{$^2$ Departament d' Estructura i Constituents de la Mat\`eria,Universitat de Barcelona.
647, Diagonal, 08028 Barcelona, Spain.}
%\affiliation{$^1$Dipartimento di Fisica dell'Universit\`a and INFN, I-16146 Genova, Italy.\\
%$^2$Departament ECM, 
%Universitat de Barcelona. 647, Diagonal, 08028 Barcelona, Spain.}

\date{\today}% It is always \today, today,

\begin{abstract}
In the framework of various statistical models as well as of mechanisms for color confinement, 
disorder parameters can be developed which are generally expressed as ratios
of partition functions and whose numerical determination is usually 
challenging. We develop an efficient method for their computation and
apply it to the study of dual
superconductivity in 4$d$ compact $U(1)$ gauge theory.
\end{abstract}

\maketitle

\section{Introduction}

Order-disorder transitions are common to a wide class of models
in Statistical Mechanics and Quantum Field Theory, the Ising
model being a prototype~\cite{Kramers:1941kn,Kadanoff:1970kz}. 
In those models, one phase is characterized by the condensation
of dual topological excitations which spontaneously breaks 
a dual symmetry, and correlation functions of 
those excitations can serve as disorder parameters for the transition:
in general they are non local in terms of the
original variables, so that their numerical study can be 
challenging. The difficulty becomes evident when the correlation
functions are expressed as ratios of partition
functions. 

Relevant examples are encountered when studying color
confinement in QCD:
that is usually believed to be related to the 
condensation of some topological excitations and models 
can be constructed accordingly, which place the
confinement-deconfinement transition into the more general scenario 
of order-disorder transitions.
One appealing model is based
on dual superconductivity of the QCD vacuum and 
relates confinement to the breaking of an abelian dual
symmetry induced by the condensation of magnetic 
monopoles~\cite{thooft75,mandelstam,parisi}.
The possibility to define disorder parameters in this scenario
has been studied since a long time~\cite{Frohlich:1986-87,fro,mar,del,DiGiacomo:1997sm}.
One parameter has been developed by the Pisa group and
is the expectation value of an operator $\mu$ 
which creates a magnetic
monopole; $\pardis$ has been shown to be a good parameter for 
confinement in $U(1)$~\cite{DiGiacomo:1997sm}, 
in pure Yang-Mills theories~\cite{PaperI,PaperIII}
and in full QCD~\cite{full1,full2}; similar parameters have been
developed both in gauge theories~\cite{moscow,bari,marchetti,bari2} 
and in statistical models~\cite{xymodel,heisenberg,ising}.
The operator $\mu$ is expressible as the
exponential of the integral 
over a time slice, $\mu = \exp(-\beta \Delta S)$ (see
later for details), so that
its expectation value can be rewritten (we consider a pure gauge theory as 
an example) as
\begin{eqnarray}
\label{defmu}
\langle \mu \rangle  = \frac{\int \left( {\cal D}U \right)  e^{-\beta (S +
\Delta S)}}{ \int \left( {\cal D}U \right)  e^{-\beta S}} = 
\frac{\int \left( {\cal D}U \right)  e^{-\beta \tilde{S}}}{ \int \left( {\cal D}U \right)  e^{-\beta S}} 
\equiv \frac{\tilde{Z}}{Z} \; ,
\end{eqnarray}
where the functional integration is over the gauge link variables, 
$S$ is the euclidean action of the theory and $\beta$ is the inverse gauge 
coupling. 
The difficulty involved in its numerical computation stems from
the poor overlap among the two statistical distributions corresponding
to the partition functions $Z$ and $\tilde Z$: configurations which
give significant contributions to $\tilde Z$ are instead extremely
rare in the original ensemble corresponding to $Z$, so that they
are very badly sampled in a Monte Carlo simulation.
The problem worsens rapidly when increasing the spatial volume, making a 
determination of $\pardis$ hardly feasible.
One way out is to evaluate 
susceptibilities of $\pardis$, like:
\beq 
\rho = \frac{d}{d \beta} \ln \pardis = 
\langle S \rangle_S -  \langle \tilde{S} \rangle_{\tilde{S}} \; , 
\label{rhodef}
\eeq
from which the disorder parameter can eventually be reconstructed as follows
\beq
%\label{rhodef}
\pardis = \exp\left({\,\int_0^{\beta} \rho(\beta^{\prime}) {\rm d}
\beta^{\prime}} \right)\; \, .
\eeq 
While that is enough to test $\pardis$ as a parameter
for confinement, 
a direct determination could be useful
in contexts like the study of its correlation 
functions~\cite{DiGiacomo:1997sm,u1mass,su2mass}.

The problem of dealing with extremely rare configurations can
be approached using the idea of generalized 
ensembles~\cite{umbrella, multicanonical, tempering}. In that framework 
we propose a new method for a direct computation of $\pardis$,
which is inspired by analogous techniques used for the
study of the 't Hooft loop~\cite{thooftloop}. We describe  
the method for the case of the $4d$ compact $U (1)$ gauge theory in the
Wilson formulation, but it is applicable to the study of 
disorder parameters in a wide class of analogous problems.

\section{The method}
\label{themethod}

The partition function of the model is defined, in the Wilson
formulation, as follows
\beq
  Z (\beta) = \int [d \theta] e^{\, - \beta S} \,
\eeq
\beq
  S = \sum_{\overrightarrow{x,} t, (\mu \nu)} \left( 1 - \cos
  (\theta_{\mu \nu} ( \vec{x}, t) \right) \,  \label{wil}.
\eeq
where the integration is over the link variables (phases in $U(1)$)
and  $\theta_{\mu \nu}( \vec{x}, t)$ is the 
plaquette in the $\mu\nu$ plane sitting at lattice site $( \vec{x}, t)$.
The model has a
critical point at $\beta_c \simeq 1.01$, which is believed to
be weak first order and separates a disordered phase ($\beta <
\beta_c$), with condensation of magnetic 
monopoles and confinement of electric charges, from a Coulomb phase
where magnetic charge condensation disappears.

The magnetically charged operator $\mu( \vec{y}, t_0)$, whose
expectation value detects dual superconductivity, 
creates a monopole 
in $\vec y$ at time $t_0$ by shifting the quantum gauge fields 
by the classical vector potential of a monopole, 
$\vec{b} ( \vec{x} - \vec{y})$,
and can be written (see Ref.~\cite{DiGiacomo:1997sm} for details) as 
\begin{equation}
  \mu ( \vec{y}, t_0) = \exp \left[ i \frac{1}{e} \int \text{\tmtextrm{d}}^3 x
  \hspace{0.25em} \vec{E} ( \vec{x}, t_0) \vec{b} ( \vec{x} - \vec{y}) \right]
  \, ,
\end{equation}
with the electric field $\vec{E} ( \vec{x}, t_0)$ being the 
momentum conjugate to the quantum vector potential.
It can be discretized on the lattice as follows: 
\beq
\mu 
= e^{\beta
   \sum_{\vec{x}, i} \left( \cos \left( \theta_{0 i} ( \vec{x}, t_0) - b_i ( \vec{x}
   - \vec{y}) \right) - \cos (\theta_{0 i} ( \vec{x}, t_0)) \right)}
\equiv
e^{-\beta \Delta S}  \; ,
\eeq
where $\theta_{0 i}$ are the phases of the temporal plaquettes,
corresponding to the electric field in the (na\"{\i}ve) continuum limit.
If we define the modified action 
\beq 
&& \tilde{S} = \sum_{\overrightarrow{x,} t \neq t_0, (\mu \nu)}
   \left( 1 - \cos (\theta_{\mu \nu} ) \right)  \nonumber \\ &+& \sum_{\vec{x}, i} \left(
   1 - \cos (\theta_{0 i} ( \vec{x}, t_0) - b_i ( \vec{x} - \vec{y})) \right)
 = S + \Delta S \;
\eeq
which differs from $S$ only on the timeslice where
the monopole has been created, we can write
\beq 
\langle \mu ( \vec{y}, t) \rangle = \tilde{Z} (\beta) / Z (\beta) 
\label{muratio} \, ; \\
\tilde{Z} (\beta) = \int [d \theta] \exp (- \beta \tilde{S} ) \, .
\eeq

Measuring $\langle \mu ( \vec{y}, t) \rangle$ in a Monte Carlo (MC) 
simulation is very difficult~\cite{DiGiacomo:1997sm}: $\mu$ gets 
significant contributions only on those 
configurations having very small statistical weight (which are poorly sampled in a finite MC
simulation). The difficulty increases with the system size as the two
distributions corresponding to $Z$ and $\tilde
Z$ shrink towards non overlapping delta functions in the configuration
space.
Our proposal is to determine the ratio in Eq.~(\ref{muratio}) by using intermediate
distribution functions having a reasonable overlap with both 
statistical ensembles corresponding to $Z$ and $\tilde Z$: our method 
has many similarities with strategies adopted in the computation of 
analogous order parameters~\cite{kovacs,rebbi,deforc,helicity}.
As a first step we rewrite the ratio as the 
product of $N$ distinct ratios:
\beq
\frac{\tilde Z}{Z} = \frac{Z_N}{Z_{N-1}} \, \frac{Z_{N-1}}{Z_{N-2}}
\, \dots \, \frac{Z_1}{Z_{0}} \; ,
\label{zratios}
\eeq
%\beq
%{\tilde Z}/{Z} = \prod_{k = 0}^{N-1} \, Z_{k+1}/{Z_{k}} 
%\label{zratios}
%\eeq
where $Z_N \equiv \tilde Z$, $Z_0 \equiv Z$ and $Z_k$ is defined in
terms of an action $S_k$ which is an interpolation between $S$ and
$\tilde S$:
\beq
Z_k \equiv \int [d \theta] e^{-\beta {S_k}} \, ; \\
S_k \equiv \frac{N - k}{N} S + \frac{k}{N} \tilde S \, \label{interpolate} .
\eeq
The idea is to compute each single ratio by a different Monte Carlo
simulation: the difficulty of dealing with $N$ simulations
should be greatly compensated by the increased overlap in the
distributions corresponding to each couple of partition functions, 
leading to a benefit which increases exponentially with $N$.
As a second step to further improve the overlap,
we compute each single ratio on the r.h.s. of
Eq.~(\ref{zratios}) using an intermediate distribution:
\beq
\frac{Z_{k+1}}{Z_k} = \frac 
{\left\langle \exp \left( - \beta \Delta {S} / 2  N \right) 
\right\rangle_{k + 1/2} }{
\left\langle \exp \left( \beta \Delta S/2 N \right) \right\rangle_{k+1/2}} \label{singleratio}\; ,
\eeq
where each expectation value is computed with the action
\beq
S_{k+1/2} \equiv \left(1 - (k + 1/2)/N \right)\, S + \left( (k + 1/2)
/ N \right) \, \tilde{S}.
\eeq
Since both expectation values in Eq.~({\ref{singleratio}) are computed
with the same MC simulation, we make use of a jackknife analysis to
get a reliable error on $Z_{k+1}/Z_k$. The final uncertainty on $\tilde Z/
Z$ is then obtained by standard error propagation since each single 
ratio on the r.h.s. of Eq.~(\ref{zratios}) is obtained by an
independent MC simulation.

Our technique of rewriting the ratio $\tilde Z/Z$ as a
product of intermediate ratios resembles very closely the
well known snake algorithm~\cite{deforc} as well as other
algorithms inspired by it, like that
used for the computation of the helicity modulus~\cite{helicity}.
However it differs from previous algorithms in the choice of the intermediate
partition functions, which in our case is not related
to the details of the model,   so that it
can  be applied without modifications
to a wider class of problems in Lattice Field Theory
and Statistical Mechanics.

Regarding the choice 
of boundary conditions (b.c.), we do it
in a consistent equal way for all the partition functions 
in Eq.~(\ref{zratios}).
We make use of both periodic and free
b.c. in the spatial directions: while one could expect a 
substantial 
difference in presence of a magnetic charge, we
will show that, in the phase where $\pardis \neq 0$, 
the two choices lead to the same thermodynamical limit; that is 
expected since in that phase the 
vacuum does not have a well defined magnetic charge and a 
monopole is completely screened.
As for  the temporal direction, 
a consistent usual choice for $\pardis$~\cite{DiGiacomo:1997sm} 
is that of periodic b.c. for $Z$ and $C^*$ b.c. for $\tilde Z$;
in particular $C^*$ boundary conditions, which corresponds to
performing a charge conjugation transformation on gauge fields
when crossing the time boundary, are taken so as to annihilate 
the monopole after one loop around the periodic time direction,
avoiding in this way that it propagates an indefinite number of times.
However we do not keep that choice, since it would lead to intermediate 
actions with inconsistent mixed b.c.; instead we adopt either free or 
periodic b.c. for both $Z$ and $\tilde Z$ also in the time direction, 
showing again that 
the choice is inessential when $\pardis \neq 0$.

\section{Numerical results}
As a first test we compare the na\"{\i}ve
computation of $\pardis$, \ie performed with a single MC
simulation using the Wilson action $S$, with our 
method for $N = 1$: we use a $4^4$ lattice
with free b.c.~at $\beta = 0.8$ and 
$10^7$ measurements for both cases,
obtaining $\pardis = 1.14(18)$ with the na\"{\i}ve computation and 
$\pardis = 0.868(3)$ with our method. Apart from the strongly reduced
error, much is learned by looking at 
the distributions of the observables $\exp(-\beta \Delta S)_S$ and 
$\exp( \pm \beta \Delta S/2)_{(S + \tilde S)/2}$ (the subscript indicates
the action used for sampling) used in the computation 
(see Eq.~\ref{singleratio}).
In Fig.~\ref{fig:histo}
we plot, for each fixed observable $O$, the distribution of the
logarithm of $O$  times the observable
itself (we choose the logarithm for graphical convenience) as a function of $\log O$, so that
the integral of each curve gives the expectation
value of the relative observable. As it is clear, for 
$\langle \exp(-\beta \Delta S) \rangle_S$ most of the contribution
comes from a region which is badly sampled: on
larger lattices the problem worsens rapidly and a na\"{\i}ve
determination of $\pardis$ is unfeasible.
The improvement obtained with our method is apparent already for $N = 1$.

In Fig.~\ref{fig:N} we show a determination of $\pardis$ for several
values of $N$ on a lattice $16^4$ with free b.c. at $\beta =
0.8$: 
for each determination a comparable whole statistics 
of $N \times N_{\rm meas} = 3.2 \cdot 10^5$ measurements has been used, so that the
error on $\pardis$ is an indication of the efficiency as a
function of $N$; in Fig.~\ref{fig:intratios} we report the intermediate
ratios $Z_{k+1}/Z_k$ (see Eq.~\ref{zratios}) used for each 
measurement.
While the intermediate ratios are strongly 
dependent on $N$, $\pardis$ is not, thus confirming the 
absence of uncontrolled systematic errors. The
statistical error rapidly changes for small values of $N$,   
but then stabilizes, indicating that a value $N \sim O(10)$
saturates the improvement.

\begin{figure}[t!]
\includegraphics*[width=\columnwidth]{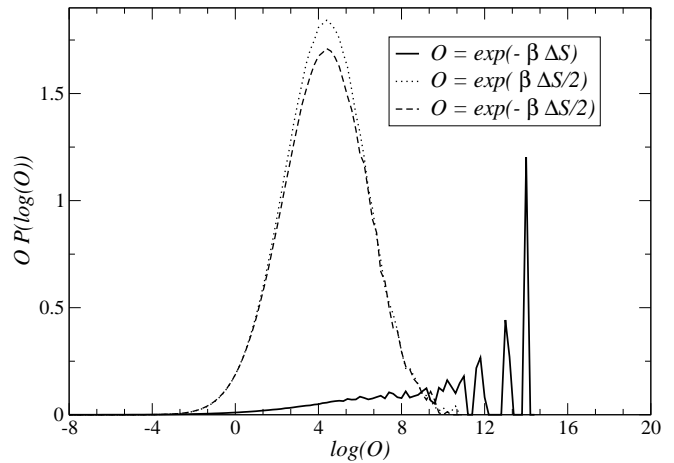}
%\vspace{0.0cm}
\caption{For each observable $O$ involved in the computation
 on the $4^4$ lattice at $\beta = 0.8$, we plot $O$ times the
 distribution
of $\log O$ as a function of $\log O$: the integral under each curve
gives $\langle O \rangle$. 
}\label{fig:histo} 
%\vspace{-0.2cm}
\end{figure}

\begin{figure}[b!]
\includegraphics*[width=\columnwidth]{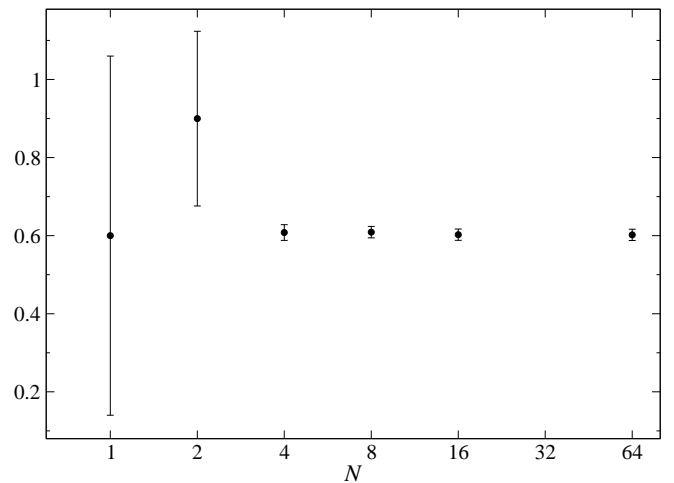}
%\vspace{-0.5cm}
\caption{Determination of $\pardis$ on a lattice $16^4$ with free
b.c. at $\beta = 0.8$ as a function of $N$.}
\label{fig:N} 
%\vspace{-0.2cm}
\end{figure}

\begin{figure}[t!]
\includegraphics*[width=\columnwidth]{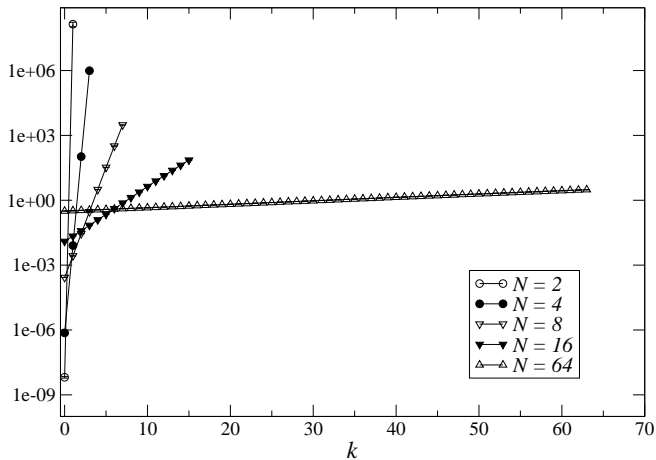}
%\vspace{-0.5cm}
\caption{Intermediate ratios $Z_{k+1}/Z_k$ used for the determinations
reported in Fig.~\ref{fig:N}.}
\label{fig:intratios} 
%\vspace{-0.2cm}
\end{figure}

\begin{figure}[b!]
\includegraphics*[width=\columnwidth]{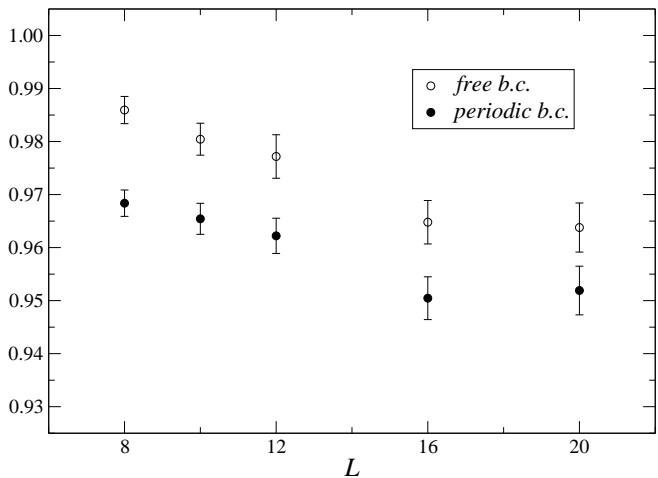}
%\vspace{-0.5cm}
\caption{$\pardis$ at $\beta = 0.5$ as a function of the lattice
size $L$ determined with both free and periodic b.c.}
\label{fig:largeL1} 
%\vspace{-0.2cm}
\end{figure}

As an application of our method we analyse some relevant features of
the disorder parameter, starting with a study of its thermodynamical limit 
in the confined phase.
In Fig.~\ref{fig:largeL1} we show $\pardis$
determined with both free and periodic b.c. at $\beta = 0.5$ as a function
of the lattice size $L$. A fit according to $\pardis = A + B/L$
gives $A = 0.945(6)$ and $B = 0.31(6)$ with free b.c. ($\tilde \chi^2 =
0.5$) and $A = 0.940(6)$ and $B = 0.2(6)$ with periodic b.c. ($\tilde \chi^2 =
0.8$). In both cases $\pardis$ has a well
defined thermodynamical limit, which does not depend (within numerical
errors) on the b.c. chosen: that is expected in
the phase where magnetic charge is 
completely screened. We stress that, contrary to what may happen 
with other parameters~\cite{helicity}, we do not expect exactly
$\pardis = 1$ in the confined phase: 
indeed any non-zero value of the
disorder parameter ensures the breaking of the magnetic symmetry, 
hence dual superconductivity.

A further confirmation of screening comes from the study of cluster
property in the correlation functions. In Table~\ref{table} we report
the values measured for various temporal and spatial correlators
of $\mu$ using periodic b.c., together with their second and fourth
powers. In this way we compare, for instance, the value of 
$\pardis^2$ (first row, second column) with that of the two point
function $\langle \bar\mu  \mu \rangle$ at large distances 
(first column, second row), or $\pardis^4$ with the four point
function, and so on: the compared quantities should approach
each other (exponentially in the extension of the higher order
correlator) if cluster property is obeyed. This is nicely verified,
within errors, from the data reported in the Table.

\begin{table}[h]
  \begin{tabular}{l|lll}
    & $\mathcal{O}$ & $\mathcal{O}^2$ & $\mathcal{O}^4$\\
    \hline
    $\langle \mu ( \vec{0}, 0) \rangle$ & 0.439(12) \  & 0.193(11)  & 0.037(4)\\
    $\langle \bar\mu ( \vec{0}, t) \mu (\vec{0}, 0) \rangle$&
    0.182(7) & 0.033(3) & \\
    $\langle \bar\mu ( \vec{z}, 0) \mu ( \vec{0}, 0) \rangle$ 
    & 0.183(12)  &  0.033(4) & \\
    $\langle \bar\mu ( \vec{z}, 0) \bar\mu ( \vec{0},
    t) \mu ( \vec{z}, t )  \mu ( \vec{0}, 0) \rangle$ & 0.037(6) &  & 
  \end{tabular}
  \caption{ Determination at $\beta = 0.8$ on a $16^4$ lattice of
   $\langle \mu \rangle$, of its spatial and temporal 2-point function 
   (second and third row) and of its mixed 4-point function (last
   row), with $t = 8$ and $\vec{z} = (0, 0, 8)$. 
   The measured correlator is reported in the first column, 
   indicated generically with $\mathcal{O}$, while in the second
   and third column (indicated with $\mathcal{O}^2$ and $\mathcal{O}^4$)
   we report respectively the second and fourth power of the
   correlator, which are used to test cluster property on the higher
   order correlators reported in lower rows.
   Within errors all measurements are compatible with the
   hypothesis that the correlators are \ already in \ their
   asymptotic regime governed by cluster property.}
\label{table}
%\vspace{-0.0cm}
\end{table}

\begin{figure}[b!]
\includegraphics*[width=\columnwidth]{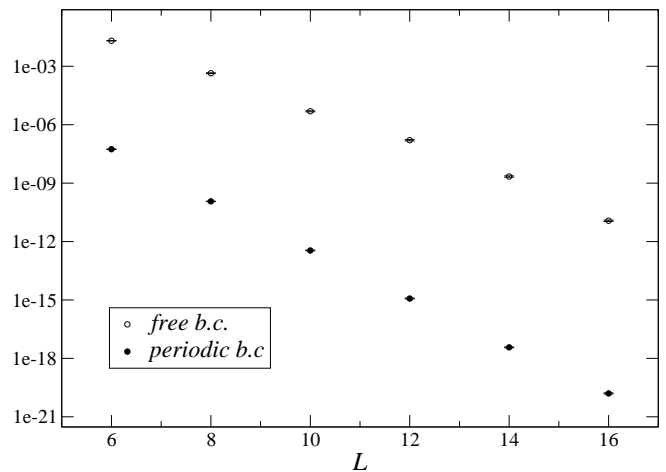}
%\vspace{-0.5cm}
\caption{$\pardis$ at $\beta = 1.1$ as a function of the lattice
size $L$ determined with both free and periodic b.c.}
\label{fig:largeL2} 
%\vspace{-0.4cm}
\end{figure}

Results are quite different in the deconfined phase. 
In Fig.~\ref{fig:largeL2} we show $\pardis$ as a function of $L$ 
at $\beta = 1.1$: the determinations with free and 
periodic b.c. differ from each other, both going to zero 
exponentially with the lattice size $L$. This is the correct expected
behavior in the phase with magnetic charge 
superselection~\cite{DiGiacomo:1997sm,supersel}.

\begin{figure}[t!]
\includegraphics*[width=1.01\columnwidth]{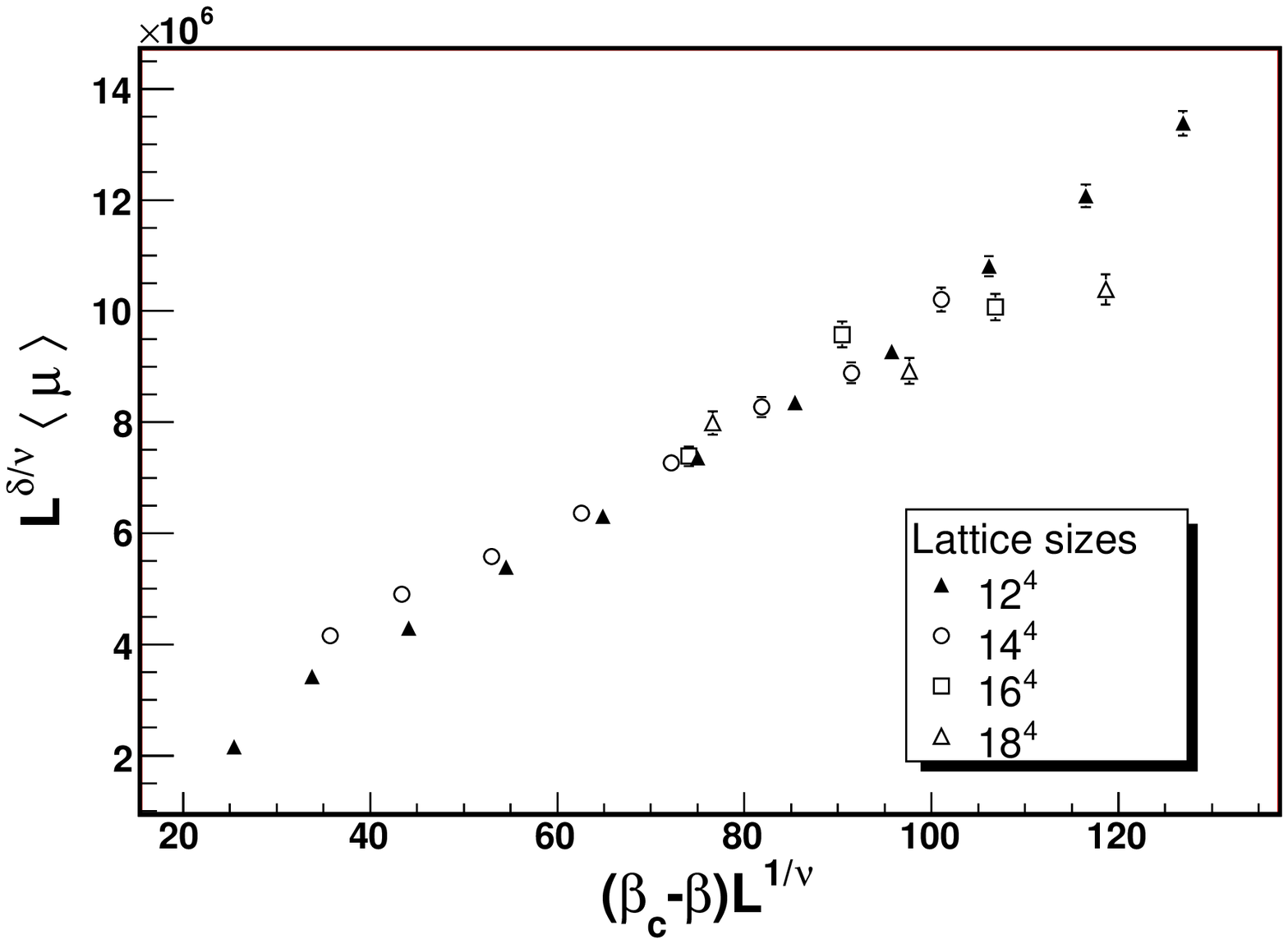}
%\vspace{-0.5cm}
\caption{F.s.s. analysis of $\pardis$ around the phase
transition}
\label{fig:fss} 
%\vspace{-0.4cm}
\end{figure}

Finally we consider  the critical behaviour of the disorder
parameter close to the phase transition, where $\pardis \simeq
\tau^\delta$ ($\tau$ is the reduced temperature). That translates in
the following finite size scaling (f.s.s.) behaviour 
\beq
\pardis = L^{- \frac{\delta}{\nu}} \phi \left( (\beta_c - \beta)
  L^{1/\nu} \right)
\eeq
where $\phi$ is a scaling function. To test this ans\"atz we have determined
$\pardis$ close to the phase transition on several different lattice sizes.
Fixing the known value of $\beta_c = 1.011$ and $\nu = 1/d = 0.25$ as
appropriate for a weak first order transition, we obtain a reasonable  
scaling with $\delta \sim 2.3$: the quality of our f.s.s. analysis
is shown in Fig.~\ref{fig:fss}.
\section{Conclusions}

We have proposed a new technique for the 
computation of disorder parameters and applied it to the study of the parameter
for dual superconductivity in 4$d$ compact U(1) 
gauge theory.  Our method is inspired by methods used for the study of the 
't Hooft loop~\cite{deforc}.

We have determined some relevant features of $\pardis$
both in the confined and in the 
Coulomb phase. A careful analysis of its critical 
properties could help in clarifying the nature of the phase transition
at zero as well as at finite temperature~\cite{finiteU1-berg,finiteU1-defor}: to that aim
also a direct comparison with analogous order parameters developed for 
$U(1)$~\cite{bari2,helicity} will be particularly useful.

Our method can be placed in the more general
framework of techniques based on the idea of generalized 
ensembles~\cite{umbrella,multicanonical,tempering}: in that respect, 
it has the advantage
to provide a recipe which can be easily applied, with none or few
modifications, to a wide class of problems. Among others we will
consider in the future  the study of order-disorder transitions in
statistical models and dual superconductivity in non Abelian gauge
theories. Another benefit of our proposal is that it leaves room 
for considerable further improvement: for instance it could be
possible to choose a more general non linear interpolation between the two 
actions $S$ and $\tilde S$, differently from what has been done in 
Eq.~(\ref{interpolate}), so as to concentrate the numerical effort
on those intermediate ensembles where the statistical distribution
is changing more rapidly.

\begin{acknowledgments}
We thank A.~Di Giacomo, A.~D'Alessandro, Ph.~de~Forcrand and 
G.~Paffuti for useful comments and discussions. L.T. thanks the
Physics Department of the University of Genova for hospitality during the
initial stages of this work.
Numerical simulations have been run on a PC farm at INFN-Genova
and on the ECM cluster of the University of  Barcelona. This work has been partially supported by MIUR.
\end{acknowledgments}

\end{document}